\documentstyle[prb,preprint,aps]{revtex}

\begin{document}
\draft

\title{Magnetic polarization  induced by nonmagnetic impurities in high Tc cuprates}
\author{Shi-Dong Liang$^{1,*}$ and T. K. Lee$^{1,2}$}
\address{$^{1}$Institute of Physics, Academia Sinica, Taiwan}
\address{$^{2}$National Center for Theoretical Sciences, Hsinchu, Taiwan}
\date{\today}


\maketitle

\begin{abstract}

The magnetic polarization induced by nonmagnetic impurities such as Zn in high Tc
cuprate compounds is studied by the variational Monte Carlo simulation.
The variational wave function is constructed from the eigenstates obtained from 
Bogoliubov de Gennes mean field Hamiltonian for the two-dimensional $t-J$ model.
A Jastrow factor is introduced to account for the induced magnetic moment
and the repulsion between holes and the impurity.
A substantial energy gain is obtained by forming an 
antiferromagnetic polarization covering 4 or 5 lattice sites 
around the impurity. We also found the doping dependence for the 
induced magnetic moment consistent with experiments.
\end{abstract}




Recently a number of  experiments, the neutron scattering\cite{sidis}, 
nuclear-magnetic-resonance (NMR)\cite{Julien,bobroff} and 
scanning tunneling microscopy (STM)\cite{pan,yazdani,hudson}, have been carried 
out to study the impurity effect on the electronic transport and magnetic properties 
in high $T_{c}$ cuprate compounds. 
These studies provide a detail information about the relationship between 
magnetism and superconductivity in high $T_{c}$ cuprates. 
The nonmagnetic impurity Zn was found to suppress $T_{c}$ more strongly than 
magnetic impurity Ni, even though both replace Cu in the CuO$_2$ 
plane\cite{mendels}. The amazingly accurate measurement of the local density of 
states (LDOS) by STM\cite{pan,yazdani,hudson} also 
provides very different spectra for Zn and Ni. 
The spin dynamics studied by the neutron scattering experiments reveals 
that the low-energy spin fluctuations are strongly enhanced near the impurity
and the magnetic excitation at the antiferromagnetic wave vector $(\pi,\pi)$ 
disappears with Zn doping in the underdoped region\cite{Kakurai,Fong}. 
It is interesting to find from the NMR and SQUID experiments that both the 
nonmagnetic Zn and the magnetic Ni impurities induce a local magnetic moment 
on Cu sites surrounding the impurity in the normal state. The broadening of $^{63}$Cu 
and $^{17}$O NMR lines has been attributed to a distribution of magnetic moments
or a spatially inhomogeneous spin polarization extending over several lattice 
sites around the impurity. On the other hand some experiments\cite{williams} 
found no evidences of the existence of local magnetic moments, at least 
in the optimum and overdoped samples. More careful theoretical and experimental
efforts to exmine the magnetic polarization are needed to clarify this issue.

So far most of the theoretical work has been based upon phenomenological 
BCS type models with emphasis on  understanding
of the LDOS. The observed nearly-zero-energy-resonance peak  near Zn impurity
was explained very early by Balatsky, Salkola and co-workers\cite{balatsky,sal1,sal2,onishi}
by assuming Zn to be an unitary impurity. 
Studies\cite{poilb,Ng,palee,khal,zhu1,tsuch} 
based upon $t-J$ type models have also successfully explained the LDOS.
There are only few studies\cite{khal} about the structure of 
magnetic polarization induced by the magnetic moment binded to 
the nonmagnetic impurity and the screening of this moment by other electrons. 
However in a recent paper\cite{tsuch2} Tsuchiura  {\it et al.} 
use Gutzwiller approximation and  the  Bogoliubov-de Gennes (BdG) approach
for the $t-J$ model and they
find no evidence of the existence of the local moments around the Zn impurity.
They also concluded that the electron avoids the impurity instead of 
being binded to it. A much more careful examination of the effect of a non-mgnetic 
impurity in  the $t-J$ model is needed to resolve the controversy.

Comparing with other phenomenological models,  the $t-J$ model has much stronger magnetic
correlation and it may lead to  a different picture about the 
magnetic polarization
around the impurity. However, previous studies of the t-J model  use
the BdG approach with or without the Gutzwiller approximation and the 
no-doubly-occupied constraint imposed by the $t-J$ model
is only taken into account on the average or approximately.
It very likely underestimates the
antiferromagnetic correlation inherent in the $t-J$ model.
Another issue has not been addressed adequately before
is the doping dependence of the induced magnetic moment. Very different 
results reported by NMR experiments\cite{Julien,mendels,williams} may
be related to the doping dependence.

In this paper we will impose the constraint rigorously by using the variational
Monte Carlo approach\cite{gros} to study the effect of 
nonmagnetic Zn impurity on the ground state of the $t-J$ model. 
The ground state trial wave function is first constructed 
by assuming d-RVB order parameters in the BdG approach. 
Then  the variational wave function is shown to be greatly improved by adding
a Jastrow factor to account for the strong magnetic correlation. 
We found a large energy gain by having an 
antiferromagnetic polarization around the impurity
with size about 4 to 5 lattice sites as observed in $^{63}$Cu NMR 
data\cite{Julien} in the underdoped region. 
The significant suppression of the magnitude of the induced
moment and its polarization size as doping increases to optimum doping 
is also consistent with experimental
observations\cite{bobroff,mendels}. In addition, our result also provides a
reason to explain the similarity between results\cite{bobroff} measured
for Li$^+$ and Zn$^{2+}$.   Contrary to the work reported
in Ref.(21) we show that electrons are always attracted to the
impurity. But the effect gets weaker when number of holes increases.

The model we consider is the dilute impurity limit of the two-dimensional
$t-J$ model. The interaction between impurities is neglected.  
Zn$^{2+}$[3d$^{10}]$ has total spin $S=0$ and its second 
ionization energy is about $18eV$. Near chemical potential 
the conduction electron  
is estimated to encounter a repulsive local potential $U_0\approx 18.9eV$\cite{zhu2} 
when it scatters with the Zn impurity.
This is much larger than the bandwidth ($2eV$) of the 
$d_{x^{2}-y^{2}}$ band of
$3d$ Cu$^{2+}$ electrons. Thus, the  nonmagnetic impurity Zn can be 
described roughly by a spin vacancy in the unitary limit.
We start from the Hamiltonian,

\begin{equation}
H=-t\sum_{<ij>,\sigma}P_{G}(c^{\dagger}_{i\sigma}c_{j\sigma}+h.c.)P_{G}
+J\sum_{<ij>}({\bf S_{i} \cdot S_{j}}-\frac{1}{4}n_{i}n_{j})
+\sum_{i}(U_0\delta_{i,I}-\mu)n_{i\sigma},
\label{H}
\end{equation}
where $I$ labels the site of the impurity. In the standard notation, 
the $<ij>$ means the summation over nearest neighbors 
and $P_{G}=\prod_{i}(1-n_{i\uparrow}n_{i\downarrow})$
is the Gutzwiller's projection operator that prohibits double occupancy. 
Within the mean field approximation, the BdG equation is derived 
\begin{equation}
\sum_{j}
\left(
\begin{array}{cc}
h_{ij}\   \  F_{ij} \\
F_{ij}^{\dagger}  -h_{ij}
\end{array}
\right)
\left(
\begin{array}{c}
u_{j}^{m} \\
v_{j}^{m}
\end{array}
\right)
=E_{m}
\left(
\begin{array}{c}
u_{i}^{m} \\
v_{i}^{m}
\end{array}
\right)
\end{equation}
where
\begin{equation}
h_{ij}=-(t\delta+\frac{1}{4}J\chi_{ij})+(U_0\delta_{I,i}-\mu)\delta_{i,j}
\label{h}
\end{equation}
\begin{equation}
F_{ij}=-\frac{1}{2}J\Delta_{ij}
\label{D}
\end{equation}
Here $u^{m}_{i}$ and $v^{m}_{i}$ are the Bogoliubov amplitudes 
corresponding to the eigenvalue $E_{m}$; $\chi_{ij}$ and $\Delta_{ij}$ 
are the bond and resonating-valence-bond (RVB) order parameters defined by 
$\chi_{ij}=\sum_{\sigma}<c^{\dagger}_{i\sigma}c_{j\sigma}>$ 
and $\Delta_{ij}=<c_{i\downarrow}c_{j\uparrow}-c_{i\uparrow}c_{j\downarrow}>$, 
respectively; $\delta$ is the hole density. They are determined self-consistentl
y by
\begin{equation}
\chi_{ij}=2\sum_{m}v^{m*}_{i}v^{m}_{j}
\label{chi}
\end{equation}
\begin{equation}
\Delta_{ij}=-2\sum_{m}u^{m*}_{i}v^{m}_{j}
\label{Delta}
\end{equation}
\begin{equation}
\delta=\frac{1}{N}\sum_{m,i}(|u^{m}_{i}|^{2}-|v^{m}_{i}|^{2}),
\label{delta}
\end{equation}
The solution found at zero temperature had already been shown 
by several groups\cite{zhu1,tsuch} to have 
a nearly-zero-energy resonance
for the LDOS when $U_0$ is very large compared to $J$ or $t$.
The order parameters $\Delta_{ij}$ near the impurity are suppressed
and a small component of s-wave pairing is induced.
In the slave-boson mean field theory,\cite{inaba} the magnetic correlation obtained is
overestimated. The simplest way to correct this deficiency is to
use the eigenvectors obtained by BdG equations to construct a variational
wave function with the projection operators rigorously imposed.
For the uniform case\cite{gros} a similar method has been used successfully.

Following the work by Yokoyama and Shiba\cite{yoko} and
Himeda {\it et al.}\cite{Ogata}, we write
this trial wave function for the ground state in terms 
of a Slater determinant of $N_{e}/2$ dimension,
\begin{equation}
|\phi>=
P_{G}(\sum_{ij}(U^{-1}V)_{ij}c^{\dagger}_{i\uparrow}c^{\dagger}_{j\downarrow})^{
N_{e}/2}|0>,
\label{phi}
\end{equation}
where $U$ and $V$ in Eq.(\ref{phi}) are the matrices of 
$u^{m}_{i}$ and $v^{m}_{j}$, respectively.  
Without the Gutzwiller projection operator, this wave function is exactly the
same as BdG ground state but with fixed $N_e$ electrons. 
The relation of this wave function with superconductivity in the absence
of impurity was discussed in Refs.(22,24). 
Most properties calculated with or without the Gutzwiller projection operator
are quite similar as shown by Zhang {\it et al.}\cite{zhang}. 
However, the spin-spin correlation calculated by BdG (Eq.(2)) 
is very much smaller than by Eq.(\ref{phi}).

It should be noted that the trial wave function in Eq.(\ref{phi}) 
is a paramagnetic RVB state without the antiferromagnetic long 
range order (AF LRO).
In a uniform system without impurity at low doping, $\delta<0.06$, this state 
is unstable\cite{TK} with respect to  the AF LRO.  
To take into account AF LRO, we could either add 
a Jastrow factor\cite{TK}, such as $\exp(-h_u\sum_{i}(-1)^{i}S^{i}_{z})$, 
to modify the trial wave function or we could include spin density wave 
order parameter\cite{chen,gia} to the original BdG equations. 
Since both approaches obtain almost identical results, we shall use 
a Jastrow factor here.

In addition to the issue of AF LRO at low doping, we  are also concerned with
the lack of consideration of strong correlation in the mean field 
theory of BdG equations. 
Use of Gutzwiller approximation\cite{tsuch2} in BdG would improve but it 
still may not be enough. 
When the no-double-occupancy constraint is included exactly, 
we could examine the issue of attraction\cite{tsuch2} 
or repulsion\cite{Ng,palee,khal} of holes by the impurity more
accurately. Hence we introduce a Jastrow factor
to reflect the influence of the impurity 
on the near-by hole distribution and magnetic polarization.
This new trial wave function is 
\begin{equation} 
|\psi_{I}>=\exp{(-\sum_{i}((-1)^{i}h_{i}S_{i}^{z}+\frac{\lambda(1-n_{i})}{R_{i}}
))}|\phi>,
\label{psi}
\end{equation}
where $R_{i}=\sqrt{(x_{i}-x_{I})^{2}+(y_{i}-y_{I})^{2}}$ is the distance 
from  the impurity site denoted by $I$. The first term in
the exponent in Eq.(\ref{psi}) 
introduces a spatial dependent staggered magnetic field, which consists of 
two terms, $h_{i}=h_{u}+\frac{h_{0}}{R_{i}}$.  $h_{u}$ 
provides a uniform AF LRO  at low doping with or without the impurity. 
$h_{0}$ is used to describe the enhanced 
AF correlation effect around the impurity. 
This enhancement will repel holes away from the impurity. Hence we include 
the second term associated with  $\lambda$ for this repulsion. Notice that if
$\lambda$ is negative, then the hole is attracted to the impurity and 
the electron is repelled from it.  
The values of  $h_{u}$, $h_{0}$ and  $\lambda$ are determined by minimizing the
variational energy.
In Eq.(\ref{psi}) we have chosen $\frac{1}{R}$ form to simulate the 
extent of the spin 
polarization around the impurity. We have examined
several other functional forms
and results are about the same as long as it covers a substantial region
around the impurity. Below we will report mostly the results obtained from
$|\psi_{I}>$  with $\frac{1}{R}$ form.  Then we will also show that
similar results are obtained 
with a different  trial function $|\psi'_{I}>$ using  
$\frac{1}{R^2}$ form. The latter has been previously shown by
G. Khaliullin {\it et al.}\cite{khal}
to be the spatial distribution of the impurity-induced moment.

Our attention is also focued on the spatial magnetic polarization near the impurity.
Without loss of generality the impurity is supposed to be situated at the center of the lattice.
Thus, we can use the periodic boundary condition for the numerical calculation.
For the 8x8, 12x12 and 16x16 lattice sizes we find that the spin cloud induced by the impurity extends
only several lattice sites and all the quantities we are
concerned with, including the local magnetization and the 
spin-spin corrlation function, have no qualitative and significant changes with 
the change of the lattice size.
This is because the lattice sizes we used are large  enough for the polarized 
spin cloud. Here we present the numerical 
results obtained for a $12\times 12$ lattice in the zero temperature limit
with $t/J=3$ and  $U_0=100J$.
In this paper $J$ is our basic energy unit.
We solve self-consistently the BdG equations and obtain the order 
parameters $\chi_{ij}$, $\Delta_{ij}$ and the BdG amplitudes $U$ and $V$. 
The pairing order parameters $\Delta_{ij}$ can be decomposed into 
extended s-wave and d-wave components as 
$\Delta_{d}(i)=\frac{1}{4}(\Delta_{x}(i)+\Delta_{-x}(i)-\Delta_{y}(i)-\Delta_{-y}(i))$ and
$\Delta_{s}(i)=\frac{1}{4}(\Delta_{x}(i)+\Delta_{-x}(i)+\Delta_{y}(i)+\Delta_{-y}(i))$.  
The d-wave component is suppressed around the impurity site 
and it induces a small s-wave pairing component which is consistent with 
other group's results\cite{tsuch,Franz}.
In principle, the Jastrow factor introduced could modify the distribution of the
order parameters.
In practice,  tuning the values of the order parameters around the 
solutions of the self-consistent BdG equations has little effect on 
the physical quantities discussed below, except a slightly lower ground state
energy is obtained.
After obtaining the BdG solution and matrices  $U$ and $V$,
we carry out the VMC simulation to determine the optimized ground state
energy.  $10^{5}$ samples were used in each MC simulation 
to measure the physical quantities. Since there are three variational
parameters: $h_u$, $h_0$ and $\lambda$, the calculation to find the
optimal solution is quite involved. Here we only report the
main results. In Fig.1 we show energy per site 
as a function of $\lambda$ for two doping concentrations. 
In Fig.1(a) for doping concentration $\delta=0.055$ and 
$h_u=0.05$, results
for $h_0=0.2$ (solid circles) and $h_0=0$  (open cirlces) 
are compared. 
Fig.1(b) shows that the lowest energy for  $\delta=0.152$ is achieved
for $h_u=0.0$, $h_0=0.0$ and $\lambda=0.2$.
It is noted that the energy is quite
sensitive to the value of $\lambda$. The lowest energy is acquired for 
positive $\lambda$, thus the hole is repelled away from the impurity  
while the moment is binded to the impurity.

We compare the optimal ground-state
energy per site calculated from the trial 
wave functions $|\phi>$ and $|\psi_{I}>$ in Table I. In the third row
we also list the total energy difference ($\Delta E$)
between these two wave functions.
The variational parameters for the  optimized wave function 
are listed in the table II.
\\
Table I: Optimal groud state energy per site as a function of hole density 
for two trial wave
functions:$|\phi>$ and $|\psi_{I}>$. The third row lists their total energy 
difference \\
\begin{tabular}{lccccccc}
\hline\hline\\
Doping $\delta$&	0.028 & 0.055 & 0.083 & 0.111 & 0.139& 0.152 \\\hline	   	
$|\phi>$& -1.207$\pm$0.002& -1.336$\pm$0.004& -1.475$\pm$0.002& -1.611$\pm$0.003
& -1.738$\pm$0.003& -1.801$\pm$0.002 \\
$|\psi_{I}>$& -1.224$\pm$0.001& -1.351$\pm$0.001& -1.487$\pm$0.002& -1.623$\pm$0.003
& -1.748$\pm$0.002& -1.806$\pm$ 0.001\\
$\Delta E$& -2.4$\pm$0.4& -2.3$\pm$0.7& -1.7$\pm$0.6& -1.7$\pm$0.9&
-1.4$\pm$0.7& -0.7$\pm$ 0.3\\\hline\hline
\end{tabular}
\\
\\
Table II: Optimized variational parameters for $|\psi_{I}>$. The values in the
parenthesis are for the clean system without impurity.\\
\begin{tabular}{lccccccc}
\hline\hline\\
Doping $\delta$&	0.028 & 0.055 & 0.083 & 0.111 & 0.139 & 0.152 \\\hline	   	
h$_u$& 0.1(0.1)& 0.05(0.05)& 0(0)& 0(0)& 0(0)& 0(0) \\
h$_{0}$& 0.3(0)& 0.2(0) & 0.05(0)& 0.03(0)& 0(0)& 0(0) \\
$\lambda$& 0.8(0)& 0.6(0) & 0.4(0)& 0.4(0)& 0.2(0)& 0.2(0) \\
\\\hline\hline
\end{tabular}
\\
\\
As shown in Table I the Jastrow factor which
simulates the magnetic
polarization around the impurity
in  Eq.(\ref{psi}) reduces 
the energy of 
the projected BdG wave function $|\phi>$ by a significant amount.
Although the energy per site has been improved only by a very 
small amount,  the total energy gain is greater than $0.7J$.
This is a very large energy gain due to
the influence of a single impurity. It also clearly demonstrates that 
BdG approach has significantly underestimated the magnetic 
correlation surrounding the impurity.

Table II shows that  $h_u$ is zero, i.e. there is no AF LRO for doping 
greater than $0.08$ with or without the impurity. This is expected as
a single impurity cannot 
induce LRO for the whole system.
At the underdoped region, for $\delta=0.083 \sim 0.11$, although there is no
uniform AF LRO, the spins around the impurity tend to form a local AF
cloud as reflected by the nonvanishing parameter $h_0$. It should be cautioned
that in this case our trial function  $|\psi_{I}>$ has broken the spin 
up-down symmetry. A more accurate description of this state should be a state
with a fluctuating local AF polarization but without fixing the moment 
in a particular direction.  For  $\delta\geq0.139$ although 
$h_0=0$ and there is no apparent magnetic polarization around the impurity,
the holes are still repelled from the impurity. This result disagrees with the
result reported by Tsuchiura {\it et al.}\cite{tsuch2}.

To examine the magnetic polarization induced around the impurity more closely, 
we have  calculated the difference of the local magnetization
$<S_{z}(R)>$  and the spin-spin correlation function 
$<S_{z}(n)S_{z}(n+R)>$ between systems with and without impurity. 
Both results are plotted in Fig.2 as a function of the square of the 
distance from
the impurity for several dopant densities.  
$(-1)^{R}(<S_{z}(R)>-<S_{z}(R)>_{0})$ shown in Fig.2(a)
indicates that $<S_{z}(R)>$ is enhanced near the impurity.
$<..>_0$ is for the clean system without impurity. 
For  $\delta\geq0.083$ there is no AF LRO and the induced magnetization
only exists within a few lattice constants around the impurity.
In Fig. 2(b) we show that the spin-spin correlation is also enhanced
near the impurity. Site $n$ is one of the nearest neighbors of the impurity.
Again the enhancement is weaker when the doping increases.
This is consistent with the experimental observation.

In Fig. (3) we plot the impurity induced  spin and charge profiles,
$\Delta S^{2}=<S^{2}_{i}>-<S^{2}_{i}>_{0}$, and 
$\Delta N_{h}=<N^{h}_{i}>-<N^{h}_{i}>_{0}$ respectively, for two different 
dopant concentrations.
Here $N^{h}_{i}=1-n_{i\sigma}-n_{i-\sigma}$. 
It can be seen that  the holes are kept away from the impurity 
and a spin cloud is formed around the impurity. As the hole doping 
increases the spin cloud becomes smaller in size.

To estimate the size of the induced magnetic polarization
and the induced moment, we calculate
$M(R)=3g<\sqrt{(\sum_{i}^{N_{R}}(-1)^{i}S^{z}_{i})^{2}}>$, 
where the Lande g factor  $g=2$ and
$N_{R}$ is the number of sites within radius
$R$ of the impurity. 
The difference between the induced magnetization with and without impurity,
$M(R)-M_{0}(R)$, is plotted as a function of $R^{2}$ in Fig.4. 
Results obtained by using  $|\phi>$ and $|\psi_{I}>$ are shown in Fig. 4(a)
and Fig. 4(b), respectively. In the inset of Fig. 4(b), results for
$\delta=0.111$, $0.139$ and $0.152$ are shown with a different scale. 
The saturation of values of $M(R)-M_{0}(R)$ at large R indicates that the
induced magnetization has a finite extent. We shall define the size of the
induced magnetic polarization to be
$R_c$. At $R=R_c$  $M(R)-M_{0}(R)$ reaches about $70\%$ of its saturated values.
This moment is much larger for
$|\psi_{I}>$ than for  $|\phi>$. Hence the local staggered magnetic field,
$h_i$, and the repulsion between impurity and hole introduced by the Jastrow
factor in Eq. (\ref{psi}) has enhanced the induced moment.

In Fig. 4(b)
we have used  $h_{i}=h_{u}+\frac{h_{0}}{R_{i}}$ and $\frac{1}{R}$ for the
repulsion between hole and impurity
in Eq.(\ref{psi}) for $|\psi_{I}>$. To examine the sensitivity of the
result to the choice of the $R$ dependence, we change $\frac{1}{R}$
to $\frac{1}{R^2}$ for both  $h_{i}$ and the repulsion term in the Jastrow 
factor. The optimized variational energies are almost the same as the 
results reported in Table I. The results for the induced magnetic polarization
is plotted in Fig. 4(c) which are quite similar to Fig. 4(b).

Results in Fig.(4) show that the in the AF LRO states or  $\delta\leq 0.083$, 
the induced magnetization is much larger.  When there is no LRO the 
induced magnetization decreases rapidly with increasing hole concentraion.  
This is consistent with experiments\cite{williams}. It is also consistent with
the theoretical result reported by
Tsuchiura  {\it et al.} \cite{tsuch2}. But we do not agree with their 
conclusion that the holes are attracted toward the impurity. 
On the contrary, we have shown above that the holes
are repelled away from the impurity to lower their kinetic energy. 
This effect might give an explanation to the similarity 
\cite{bobroff} between
Li$^+$ and Zn$^{2+}$. The holes are also repelled away from the Li 
to gain  energy.

The induced moment $M=(M(R_c)-M_{0}(R_c))/N_{R_c}$ and the square of the
size of the induced cloud, $R_c^2$ are plotted  as a function of hole
concentration in Fig. 5 and its inset, respectively.
For the hole concentration $\delta\leq0.055$, 
the local magnetic moment we obtained is about $0.5 \mu_B$ 
as compared with the
experimental value $0.4\sim 1$ $\mu_B$ for the Zn 4$\%$ substitution and different dopings. 
The rapid decrease of the size of the induced spin cloud 
could be due to the screening by the
conducting carriers\cite{palee,bobroff}.

In summary, the magnetic polarization induced by nonmagnetic impurities in high Tc
cuprate compounds is studied by combining
the variational Monte Carlo simulation and 
Bogoliubov de Gennes mean field Hamiltonian for the two-dimensional $t-J$ model.
A Jastrow factor is introduced to account for the induced magnetic moment
and the repulsion between holes and the impurity.
A substantial energy gain is obtained when the holes are repelled and
the antiferromagnetic polarization is enhanced near the impurity.
The doping dependence for the 
induced magnetic moment is consistent with experiments.

The author Liang would like to thank C.S Ting, J.X. Zhu, J.X. Li, 
and Chi-Ho Cheng for helpful discussions. This work is supported by 
the grant NSC89-2112-M-001-103.

* Present address:   Department of Physics, Zhongshan University, Guangzhou, 510275, PR China.

\newpage 
\begin{figure}[tbp]
\caption{Variational energies plotted as a function of the imurity-hole
repulsion parameter $\lambda$. Fig. (a) and (b) are for different values of
the parameters as speicifed in the figure.}
\end{figure}

\begin{figure}[tbp]
\caption{Enhancement  of (a) $S_{z}$ and (b) 
spin-spin correlation function  for diffferent hole densities plotted as a 
function of the square of the distance from the impurity.}
\end{figure}

\begin{figure}[tbp]
\caption{Spin and charge profiles for different hole densities
calculated by $|\psi_{I}>$.
The parameters are listed in the table II.}
\end{figure}

\begin{figure}[tbp]
\caption{The induced magnetic polarization, $M(R)-M_{0}(R)$, 
plotted as a function of $R^2$ for different  hole densities, obtained from
(a) the BdG wave function $|\phi>$ and (b) the wave function 
$|\psi_{I}>$. The parameters are listed in the table II.
In the inset, $\delta=0.111$, $0.139$ and $0.152$ are shown with a different scale. 
(c) is obtained from $|\psi'_{I}>$ which is
similar to  $|\psi_{I}>$ used in (b) but with a different functional
form for the Jastrow 
factor as discussed in the text. 
The parameters used
are $\delta=0.028$, $h_{u}=0.1$, $h_{0}=0.2$, $\lambda=1.2$; 
$\delta=0.055$, $h_{u}=0.05$, $h_{0}=0.1$, $\lambda=1.0$;
$\delta=0.083$, $h_{u}=0$, $h_{0}=0.025$, $\lambda=0.8$.}
\end{figure}

\begin{figure}[tbp]
\caption{The induced magnetic moment plotted as
a function of hole density. In the inset,
the size of the spin cloud versus hole density. The parameters are listed in 
Table II. }
\end{figure}

\newpage

\begin{figure}
\vbox to7.0in{\rule{0pt}{7.0in}}
\includegraphics{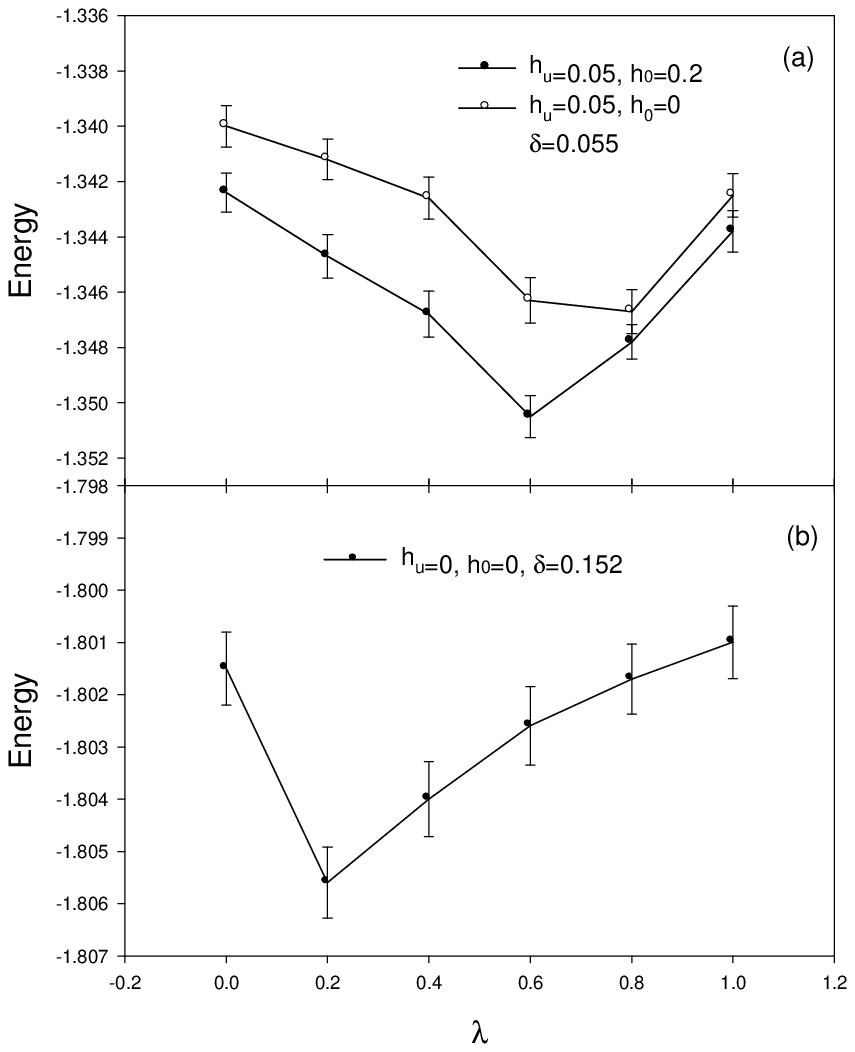}
\end{figure}

\newpage
\begin{figure}
\vbox to7.0in{\rule{0pt}{7.0in}}
\includegraphics{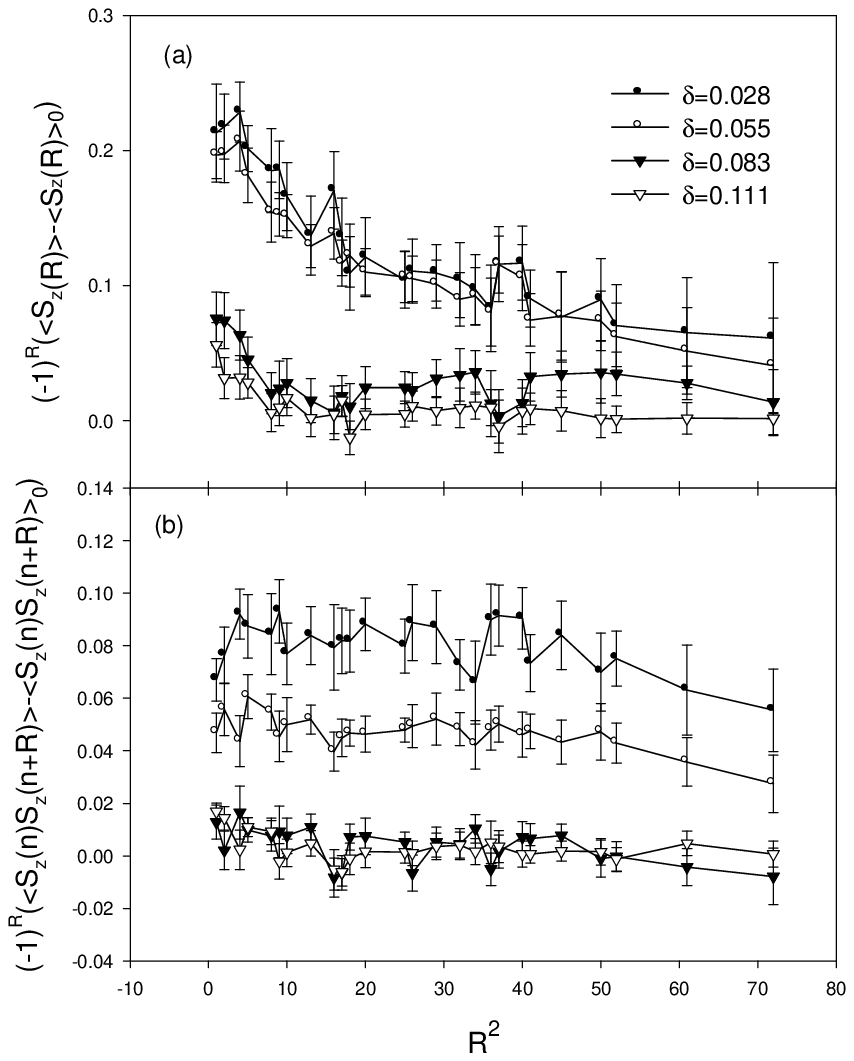}
\end{figure}

\begin{figure}
\vbox to7.0in{\rule{0pt}{7.0in}}
\includegraphics{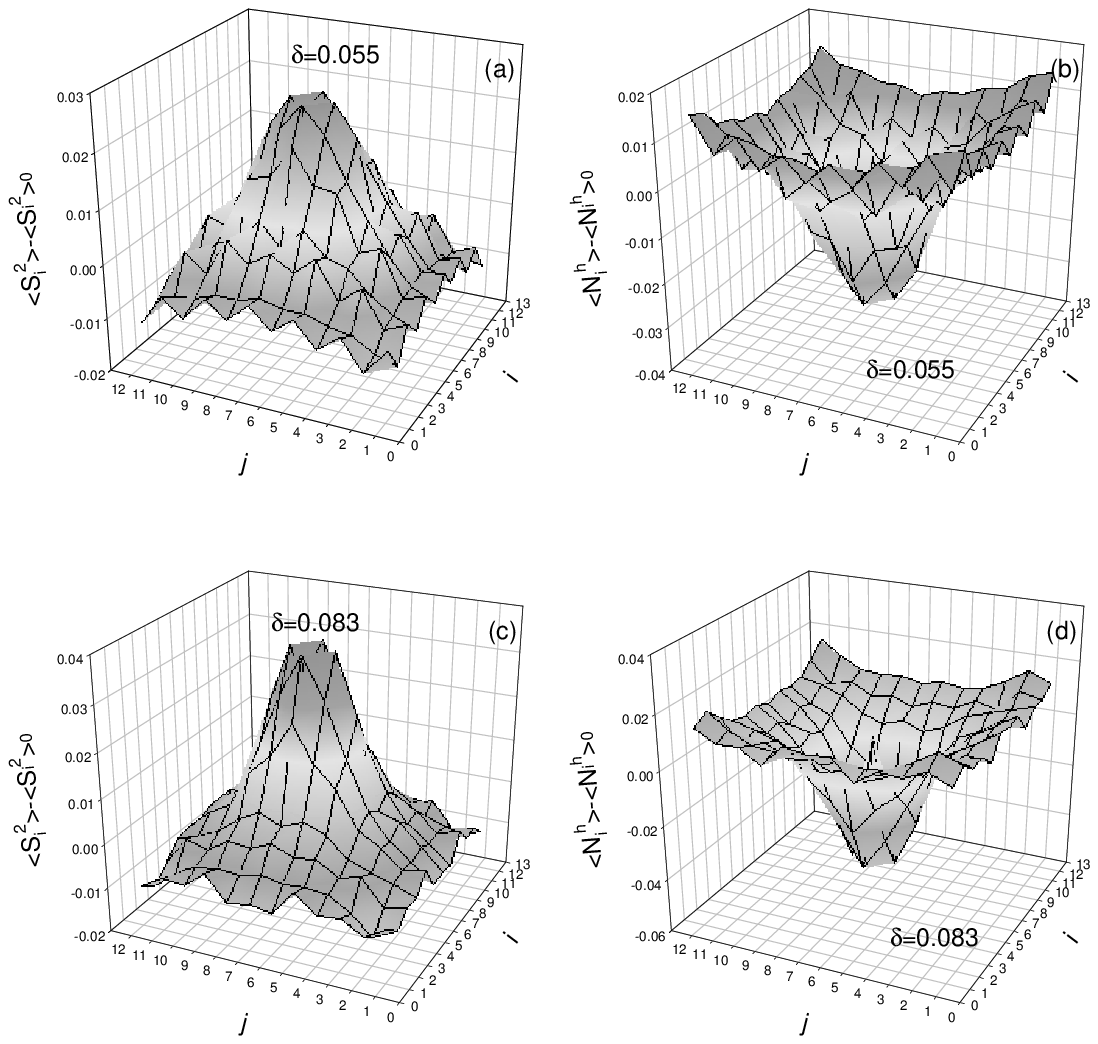}
\end{figure}

\begin{figure}
\vbox to10.0in{\rule{0pt}{10.0in}}
\includegraphics{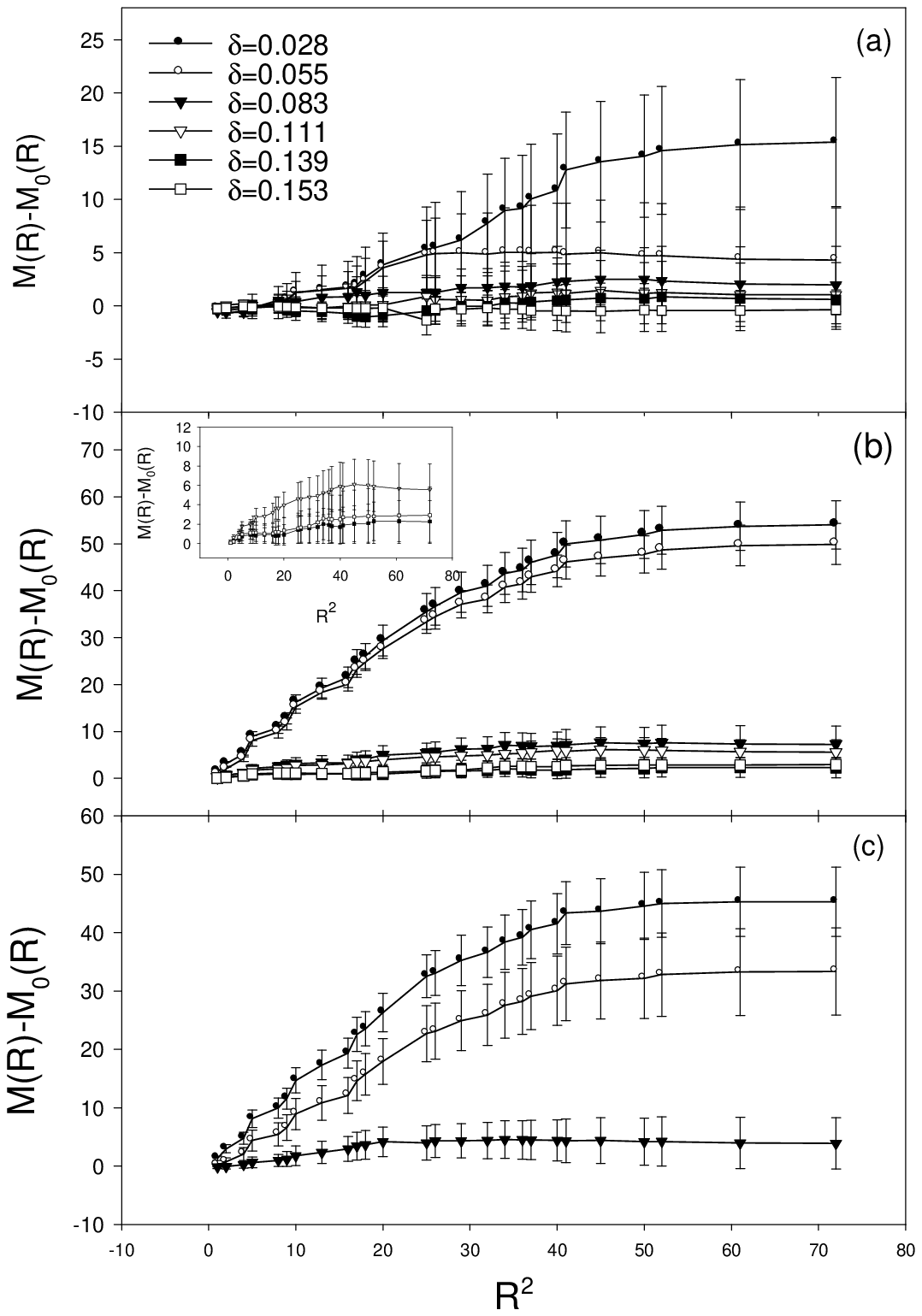}
\end{figure}

\begin{figure}
\vbox to7.0in{\rule{0pt}{7.0in}}
\includegraphics{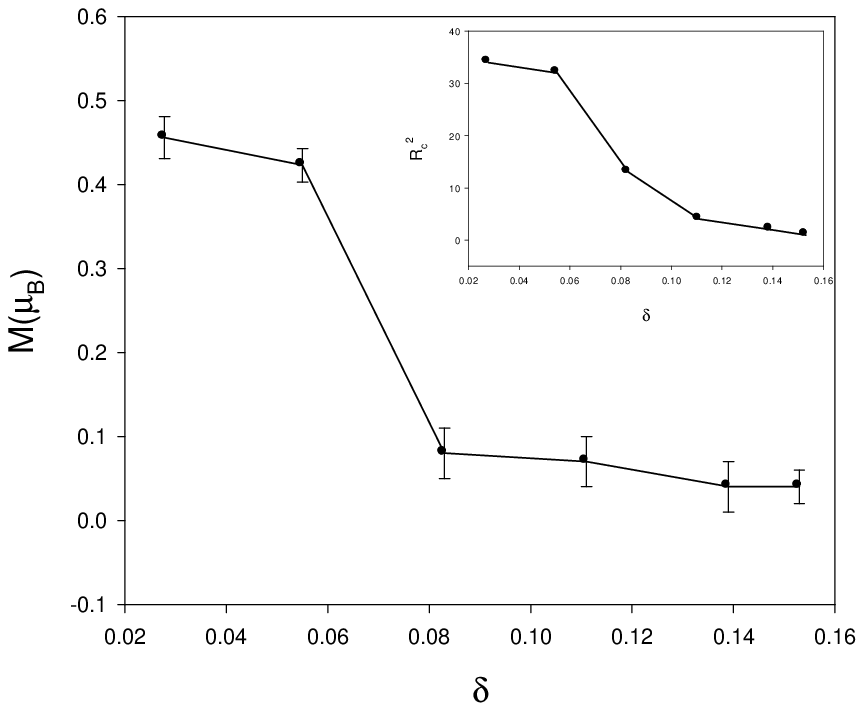}
\end{figure}


\begin{references}
\bibitem{sidis} Y. Sidis, P. Bourges, H. F. Fong, B. Keimer, L. P. Regnault, J. Bossy, 
A,Ivanov, B. Hennion, P. Gautier-Picard, G. Collin, D. L. Millius and I. A. Aksay, Phys. Rev. Lett. {\bf84}, 5900 (2000).
\bibitem{Julien} M.H. Julien, T. Feher, M. Horvatic, C. Berthier, O. N. Bakharev, P. Segransan,
G. Collin and J. F. Marucco, Phys. Rev. Lett.{\bf 84}, 3422 (2000).
\bibitem{bobroff} J. Bobroff, W. A. MacFarlane, H. Alloul, P. Mendels, N. Blanchard, G. Collin
and J. F. Marucco, Phys. Rev. Lett. {\bf 83}, 4381 (1999).
\bibitem{pan}  S.H. Pan, E. W. Hudson, K. M. Lang, H. Eisaki, S. Uchida and J. C. Davis, Nature, {\bf 403}, 746 (2000). 
\bibitem{yazdani} Ali. Yazdani, C. M. Howald, C. P. Lutz, A Kapitulnik and D. M. Eigler, Phys. Rev. Lett. {\bf 83}, 176 (1999).
\bibitem{hudson} E. W. Hudson, K. M. Lang, V. Madhavan, S.H. Pan, H. Eisaki, S. Uchida and J. C. Davis, Nature, {\bf 411}, 920 (2001).
\bibitem{mendels} P. Mendels, J. Bobroff, G. Collin, H. Alloul, M. Gabay, J. F. Marucco, N. Blanchard  
and B. Grenier, Europhys. lett. {\bf 46}, 678(1999).
\bibitem{Kakurai} K. Kakurai, S. Shamoto, T. Kiyokura, M. Sato, J, M. Tranquada and G. Shirane, Phys. Rev. B{\bf 48}, 3485 (1993).
\bibitem{Fong} H. F. Fong, P. Bourges, Y. Sidis, L. P. Regnault, J. Bossy, 
A,Ivanov, D. L. Millius I. A. Aksay and  B. Keimer, , Phys. Rev. Lett. {\bf 82}, 1939 (1999).
\bibitem{williams} G. V. M. Williams, J. L. Tallon and R. Dupree, Phys. Rev. B{\bf 61}, 4319 (2000).
\bibitem{balatsky} A. V. Balatsky, M. I. Salkola and A. Rosengren, 
Phys. Rev. B{\bf51}, 15547 (1995).
\bibitem{sal1} M. I. Salkola, A. V. Balatsky and D. J. Scalapino, 
Phys. Rev. Lett. {\bf 77}, 1841 (1996).
\bibitem{sal2} M.I. Salkola, A.V. Balatsky, and J.R. Schrieffer, 
Phys. Rev. B{\bf 55}, 12648 (1997).
\bibitem{onishi} Yoshifumi Onishi, Yoji Ohashi, Yasunori Shingki and Kazumasa Miyake, J. Phys. Soc. Jpn {\bf 65}, 675 (1996).
\bibitem{poilb} D. Poilblanc, D.J. Scalapino, and W. Hanke, Phys. Rev. Lett. 
{\bf 72}, 884, (1994); Phys. Rev. B{\bf 50}, 13020(1994).
\bibitem{Ng} Naoto Nagaosa and Tai-Kai Ng, Phys. Rev. B {\bf 51}, 15588(1995).
\bibitem{palee}Naoto Nagaosa and Patrick A. Lee, Phys. Rev. Lett. {\bf 79}, 3755
(1997); Patrick A. Lee, Phys. Rev. Lett. {\bf 71}, 1887(1993).
\bibitem{khal} G Khaliullin, R. Kilian, S. Krivenko, and P. Fulde, 
Phys. Rev. B{\bf 56}, 11882 (1997).
\bibitem{zhu1} Jian-Xin Zhu, T.K. Lee, C.S. Ting and C.R. Hu, Phys. Rev. B{\bf 61}, 8667 (2000);
Jian-Xin Zhu, C.S. Ting, and Chia-Ren Hu, Phys. Rev.B{\bf 62}, 6027(2000).
\bibitem{tsuch} Hiroki Tsuchiura, Yukio Tanaka, Masao Ogata and Satoshi Kashiwaya, J. Phys. Soc. Jpn {\bf 68}, 2510
(1999); Phys. Rev. Lett. {\bf 84}, 3165 (2000).
\bibitem{tsuch2}Hiroki Tsuchiura, Yukio Tanaka, Masao Ogata and Satoshi Kashiwaya , Phys. Rev. B{\bf 64}, 140501(R) (2001).
\bibitem{inaba} M. Inaba, H. Matsukawa, M. Saitoh and H. Fukuyama, Physica {\bf C257},299 (1996).
\bibitem{gros} C. Gros, Phys. Rev. B{\bf 38},931 (1988).
\bibitem{zhu2} Jian-Xin Zhu and C.S. Ting, Phys. Rev. B{\bf 64}, 060501(2001).
\bibitem{yoko} H. Yokoyama and H. Shiba, J. Phys. Soc. Jpn {\bf 57}, 2482 (1988).
\bibitem{Ogata} A. Himeda, T. Kato and M. Ogata, Phys. Rev. Lett. {\bf 88}, 117001(2002).
\bibitem{zhang} F. C. Zhang, C. Gros, T. M. Rice and H. Shiba,Supercond. Sci. Technol. {\bf 1},36 (1988).
\bibitem{TK}T.K. Lee and Shiping Feng, Phys. Rev. B{\bf 38}, 11809(1988).
\bibitem{chen} G. J. Chen, Robert Joynt, F. C. Zhang and C. Gros, Phys. Rev. B{\bf 42}, 2662 (1990).
\bibitem{gia} T. Giamarchi and C. Lhuillier, Phys. Rev. B{\bf 43}, 12943 (1991).
\bibitem{Franz} M. Franz, C. Kallin and A.J. Berlinsky, Phys. Rev. B{\bf 54}, R6897(1996); 
D.J. Thouless, Phys. Rev. B{\bf 13}, 93(1974).
\end{references}
\end{document}